\documentclass[pra,twocolumn,showpacs]{revtex4}
\usepackage{graphicx}
\begin{document}

\bibliographystyle{unsrt}

\title{Time-bin entangled photon holes}
\author{J. Liang}
\altaffiliation[Present address: ]{Optical Nanostructures Laboratory, Columbia University, New York, New York, 10027}
\author{J.D. Franson}
\author{T.B. Pittman}
\affiliation{Physics Department, University of Maryland Baltimore
County, Baltimore, MD 21250}

\begin{abstract}
The general concept of entangled photon holes is based on a correlated absence of photon pairs in an otherwise constant optical background. Here we consider the specialized case when this background is confined to two well-defined time bins, which allows the formation of time-bin entangled photon holes. We show that when the typical coherent state background is replaced by a true single-photon (Fock state) background, the basic time-bin entangled photon-hole state becomes equivalent to one of the time-bin entangled photon-pair states. We experimentally demonstrate these ideas using a parametric down-conversion photon-pair source, linear optics, and post-selection to violate a Bell inequality with time-bin entangled photon holes.
\end{abstract}

\pacs{03.67.Bg, 42.50.Dv, 42.50.St, 42.65.Lm}

\maketitle

\section{Introduction}
\label{sec:introduction}

Entangled photon hole (EPH) states are new form of entanglement that is based on the existence of ``missing pairs'' of photons in two optical modes \cite{franson06}. These states have now been observed in the laboratory \cite{pittman06}, and their nonclassical properties have been experimentally demonstrated \cite{afek10}. It has recently been shown that these states can be relatively insensitive to loss and amplification noise \cite{franson11}, which may be beneficial for quantum communication applications such as quantum key distribution \cite{ekert05}.

All of these initial studies \cite{franson06,pittman06,afek10,franson11,rosenblum10} have only considered a kind of {\em energy-time} EPH state \cite{franson89} where, roughly speaking, the time-scale associated with the correlated absences (ie. the ``photon-holes'') is much shorter than the coherence-time of the background states in which they exist. Motivated by the many different forms of entangled photon-pair states (eg. polarization entanglement \cite{shih88,kwiat95}, momentum entanglement \cite{rarity90}, etc.), this naturally raises the question of what other forms of EPH states might look like. Along these lines, in this paper we introduce the idea of {\em time-bin} EPH states, in analogy with time-bin entangled photon-pair states pioneered by Gisin's group \cite{brendel99,tittel00}.

The initial EPH studies \cite{franson06,pittman06,afek10,franson11,rosenblum10} have also only considered the case where the background in which the photon-holes exist is formed by coherent states. This raises questions about the possibility of photon-holes in different types of background states, such as thermal states or even various nonclassical states. Here we consider time-bin EPH's in a background formed by highly nonclassical single-photon states.

The combination of time-bins and single-photon background is particularly useful for illustrating the concept of EPH's by comparison and contrast with more familiar entangled photon-pair states. Indeed, as the coherent state background is replaced by single-photon state background, a basic time-bin EPH state becomes fully equivalent to an antisymmetric time-bin entangled photon-pair state.  This relatively simple scenario provides insight into the properties of more general EPH states, such as energy-time EPH's in coherent-state background \cite{franson06,franson11}, which do not have direct equivalences with entangled photon-pair states.

In section \ref{sec:overview}, we overview the basic idea of time-bin EPH's and transition from a general coherent-state background to a background formed formed by single-photon states. In section \ref{sec:experiment} we describe a proof-of-principle experimental demonstration of time-bin EPH's in single-photon background using a conventional parametric down-conversion (PDC) photon-pair source \cite{liangPhD12} and techniques from the Linear Optics Quantum Computing (LOQC) toolbox \cite{knill01}. Here the time-bin EPH states are sent into a ``Franson interferometer'' \cite{franson89} and used to violate a Bell-type inequality \cite{bellbook}. In section \ref{sec:summary} we summarize and discuss further applications of EPH states.

\section{Overview}
\label{sec:overview}

Perhaps the easiest way to visualize EPH states is by contrasting them with well-known photon-pair states produced by PDC.  The original proposal for energy-time EPH's used two-photon absorption as a nonlinearity which simultaneously removes one photon from each of two coherent states \cite{franson06}. In some sense, this is the direct opposite of continuous-wave (cw) pumped PDC, where a $\chi^{(2)}$ nonlinearity is used to simultaneously create one photon in each of two vacuum states \cite{klyshkobook}. In both cases, the creation/removal of the two photons occurs at the same time, but that time is completely uncertain, which is essentially the origin of the energy-time entanglement in these states \cite{franson89}.

In this way, EPH's can be thought of as the ``negative image'' of PDC \cite{pittman06}. Yet this analogy is not complete because the idea of a ``single hole'' (in contrast to a single PDC photon) is meaningless, and there can be many ``missing pairs'' in an EPH state, whereas an idealized PDC state contains just one photon-pair. Consequently, a better analogy is drawn by contrasting the relevant two-photon amplitudes involved in PDC vs. EPH states \cite{franson06}.

\begin{figure}[t]
\includegraphics[width=3.5in]{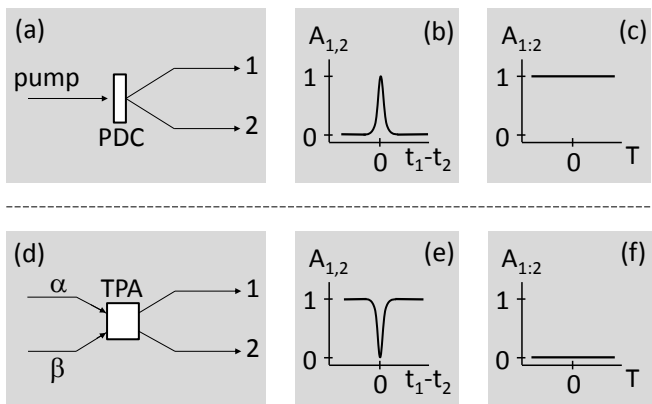}
\caption{Pictorial representation contrasting {\em energy-time} entangled PDC photon-pair states ((a)-(c)), with {\em energy-time} entangled photon-hole (EPH) states ((d)-(f)) realized through strong two-photon absorption (TPA)\protect\cite{franson06}. By contrasting the two-photon amplitudes in (b) vs. (e), as well as (c) vs. (f), it can be seen that EPH's can be thought of as the ``negative image'' of PDC. A description of the two kinds of two-photon amplitudes, $A_{1,2}$ and $A_{1:2}$, is found in the main text. The amplitudes have been normalized to 1 in each plot for convenience.}
\label{fig:energytimeoverview}
\end{figure}

To illustrate this analogy, Figure \ref{fig:energytimeoverview} shows two different forms of two-photon amplitudes (ie. the biphoton picture \cite{shihbook}) for both energy-time PDC states and energy-time EPH states. Here the term $A_{1:2}$ represents the probability amplitude for detecting one photon in each output mode ($1$ and $2$) at the same time, as function of an overall time $T$. The term $A_{1,2}$ represents the probability amplitude for detecting two photons at two different times ($t_{1}$ and $t_{2}$), as a function of the difference in these times.

Figure \ref{fig:energytimeoverview}(a) shows the familiar scenario for an idealized cw PDC source. Here
the PDC crystal can only produce the two photons at the same time, so the amplitude $A_{1,2}$ is zero everywhere except near $t_{1}-t_{2}=0$ as shown in Figure \ref{fig:energytimeoverview}(b)  (the width and shape of this peak is determined by phase-matching considerations and other details of the PDC process \cite{klyshkobook}). Due to the cw nature of the PDC pump, this photon-pair can be produced at any time with equal probability, so the amplitude $A_{1:2}$ is a constant value for all times $T$ as shown in Figure \ref{fig:energytimeoverview}(c).

In contrast, Figure \ref{fig:energytimeoverview}(d) shows the energy-time EPH scenario of reference \cite{franson06}. Here, two narrowband weak coherent-states $|\alpha\rangle$ and $|\beta\rangle$ pass through an idealized two-photon absorption (TPA) medium. The central frequencies of these coherent-states are non-degenerate, and chosen in such a way that the TPA medium does nothing to either beam by itself, but absorbs one photon from each state if they pass through at the same time \cite{franson06}. Consequently, the amplitude $A_{1,2}$ is a constant value everywhere, but goes towards zero near $t_{1}-t_{2}=0$ as shown in Figure \ref{fig:energytimeoverview}(e)(the width and shape of this dip is determined by the lifetimes of the relevant states and other properties of the TPA system \cite{boydbook}). In  the the idealized limit of infinitely strong TPA, the probability of detecting two photons at the same time is always zero, so the amplitude $A_{1:2} ~ 0$  for all times $T$ as shown in Figure \ref{fig:energytimeoverview}(f).

Contrasting the two-photon amplitudes of Figures \ref{fig:energytimeoverview}(b) vs. \ref{fig:energytimeoverview}(e), and Figures \ref{fig:energytimeoverview}(c) vs. \ref{fig:energytimeoverview}(f) provides a clear analogy of EPH's as the ``negative image'' of PDC.

We now use this analogy to introduce the idea of time-bin EPH's, which is the primary thrust of this paper. The main idea is to confine the usual cw background of Figure \ref{fig:energytimeoverview}(d) into two well-defined time-bins. As in the case of Gisin's time-bin entangled photon pairs \cite{brendel99,tittel00}, this is accomplished using short optical pulses and unbalanced Mach-Zehnder interferometers (MZ's).

\begin{figure}[b]
\includegraphics[width=3.5in]{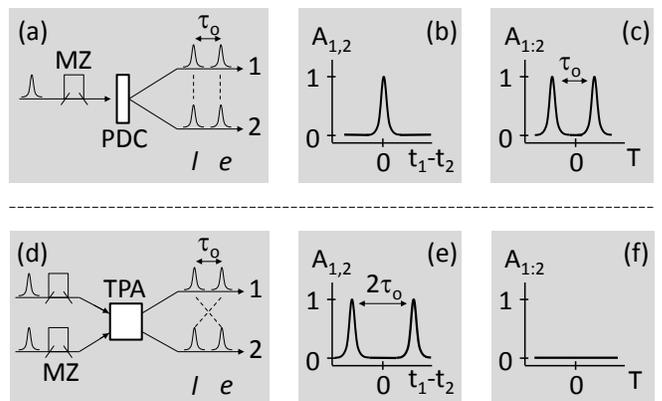}
\caption{Pictorial representation contrasting {\em time-bin} entangled PDC states ((a)-(c)) realized through the use of a pump Mach-Zehnder interferometer (MZ) \protect\cite{brendel99,tittel00}, with {\em time-bin} EPH states ((d)-(f)). The introduction of these time-bin EPH states is the main thrust of this paper. The optical pulses on the left side of (d) form the background in which the time-bin EPH's will exist. These pulses can be either coherent-state pulses or nonclassical single-photon pulses as described in the text.}
\label{fig:timebinoverview}
\end{figure}

Figure \ref{fig:timebinoverview}(a) first shows the standard scenario for time-bin entangled photon pairs produced by pulsed PDC \cite{brendel99,tittel00}. Here a short pumping pulse is divided into two coherent amplitudes by a MZ before the PDC crystal. The source produces one pair of photons in a superposition of two time bins (labeled $e$ and $l$ for ``early'' and ``late'') separated by the MZ imbalance time $\tau_{o}$. The photons will always be found in the same time-bin, so again $A_{1,2}$ is only non-zero when $t_{1}-t_{2}=0$ as shown in Figure \ref{fig:timebinoverview}(b). However, the possible pair production time is now confined to the two time bins, so $A_{1:2}$ shows only two peaks separated by $\tau_{o}$ as shown in Figure \ref{fig:timebinoverview}(c).

Time-bin EPH states are now constructed by designing a system that produces the ``negative image'' of Figures \ref{fig:timebinoverview}(b) and \ref{fig:timebinoverview}(c). One conceptual method is illustrated in Figure \ref{fig:timebinoverview}(d). Here two non-degenerate coherent-state pulses pass through two independent MZ's before entering the idealized TPA medium. The response time of the TPA process is assumed to be much shorter than $\tau_{o}$, so that pairs of photons (one in each mode) in different time bins pass through without absorption. However, the possibility of two photons (one in each mode) exiting in the same time-bin is removed by the TPA process. Consequently, $A_{1,2}$ shows two separated background peaks, but a missing central peak at $t_{1}-t_{2}=0$ as shown in Figure \ref{fig:timebinoverview}(e). The TPA process also causes $A_{1:2}=0$ in all time-bins, as shown \ref{fig:timebinoverview}(f).

Considering the relevant time-bins, by contrasting the amplitudes shown in Figures \ref{fig:timebinoverview}(b) vs. \ref{fig:timebinoverview}(e), and Figures \ref{fig:timebinoverview}(c) vs. \ref{fig:timebinoverview}(f), it can be seen that the pulsed source of Figure \ref{fig:timebinoverview}(d) generates time-bin EPH's.

It is important to emphasize that the pulses of Figure \ref{fig:timebinoverview}(d) are thus far assumed to be coherent states, which provides time-bin EPH's in a coherent state background. This case is particularly relevant, as it is essentially the coherent state background which allows EPH states to be amplified under certain conditions \cite{franson11}, and time-bin techniques are known to be generally robust in quantum systems \cite{tittel00}. Because the dominant sources of errors in quantum communication systems can originate from loss and polarization errors \cite{gisin02}, the combination of these two ideas might offer practical benefits for future systems.

On a more fundamental level, however, it is interesting to consider replacing the coherent state input pulses in Figure \ref{fig:timebinoverview}(d) with single-photon input pules. In this case, the two-photon amplitudes in Figures \ref{fig:timebinoverview}(e) and \ref{fig:timebinoverview}(f) remain the same, but the system produces time-bin EPH's in a single-photon background. This case is particularly instructive, because a time-bin EPH state in a background formed by single-photon states becomes physically identical to one of the conventional time-bin entangled photon-pair Bell-states:

\begin{equation}
|\psi^{-}\rangle=\frac{1}{\sqrt{2}}\left(|l\rangle_{1}|e\rangle_{2} - |e\rangle_{1}|l\rangle_{2}\right)
\label{eq:ELvsLE}
\end{equation}

\noindent where the kets $|e\rangle$ and $|l\rangle$ represent, respectively, a single-photon in the early or late time-bin.

In other words, there are known to be exactly two photons in the system (one in each output mode), but the existence of the holes prevents them from being found in the same time-bin. This appears to be the limiting case where EPH states and entangled photon-pair states reduce to the same physical state. Therefore, this relatively simple scenario provides insight into the fundamental nature of EPH states in more complicated scenarios such as energy-time EPH's in a single-photon backgrounds, and more useful EPH states in coherent-state backgrounds \cite{franson11}.

Additionally, this unique combination of time-bins and single-photon background allows an experimental investigation of time-bin EPH's with current technology. Indeed, a direct realization of the system sketched in Figure \ref{fig:timebinoverview}(d) (for coherent-state or single-photon backgrounds) would require strong TPA at single-photon intensities, which is exactly the type of nonlinearity needed for photonic quantum computing \cite{milburn89,franson04} and is beyond the reach of current technology. However, by considering single-photon backgrounds, proof-of principle EPH experiments can be done by starting with two single photons and putting them in time-bins in such a way as to realize the state of eq. (\ref{eq:ELvsLE}). In our experiment, we accomplish this using two photons produced from a typical cw PDC source, as will be described in the next section.

\section{Experiment}
\label{sec:experiment}

\begin{figure}[b]
\includegraphics[width=3.5in]{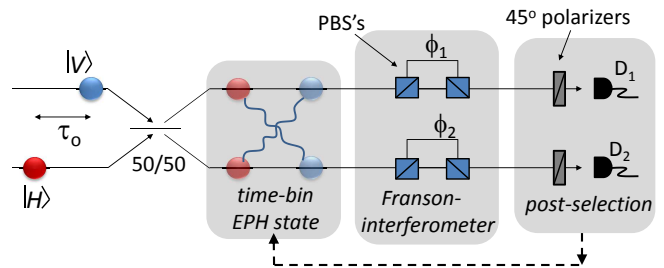}
\caption{(color online) Schematic overview of our experimental method to violate Bell's inequality using time-bin EPH's in a single-photon background. The background is formed by two single-photons that are displaced by $\tau{o}$ and mixed at a 50/50 beamsplitter. Post-selection at the output of a Franson-interferometer realizes the time-bin EPH state in the background state given by eq. (\protect\ref{eq:SLvsLS}). Polarization-encoding and polarizing beamsplitters (PBS) are used to simplify the actual experiment \protect\cite{strekalov96}.}
\label{fig:experimentoverview}
\end{figure}

A conceptual overview of our experimental demonstration of time-bin EPH's is shown in Figure \ref{fig:experimentoverview}. The idea is to perform a Bell-inequality test by sending time-bin EPH states through a conventional Franson interferometer \cite{franson89}. For simplicity, we consider a Franson arrangement using the polarization-based unbalanced MZ interferometers of Reference \cite{strekalov96} to eliminate the need for large path-length imbalances and fast electronics.

 The EPH state of interest is generated by forming the background using two orthogonally polarized single-photons (denoted by the red and blue circles labeled $|H\rangle$ and $|V\rangle$) separated by a time $\tau_{o}$, and mixing them at a 50/50 beamsplitter. Because post-selection in the output of the Franson interferometer will only register cases in which one background photon passes through each MZ, the relevant part of the two-photon state at the output of the 50/50 beamsplitter is given by eq. (\ref{eq:ELvsLE}). The  ``early-late'' and ``late-early'' terms of this background photon-pair state are denoted graphically by the two ``linked'' amplitudes in Figure \ref{fig:experimentoverview}, while the EPH state itself is perhaps best described by the two-photon amplitudes of Figures \ref{fig:timebinoverview}(e) and \ref{fig:timebinoverview}(f).

The polarizing beamsplitters (PBS's) direct horizontally polarized amplitudes along the short arms of the MZ's, and vertically polarized amplitudes along the long arms, while polarizers oriented at $45^{o}$ in the outputs erase any which-path information \cite{strekalov96}. Consequently, the Franson interferometer of Figure \ref{fig:experimentoverview} transforms the background photon-pair state of eq. (\ref{eq:ELvsLE}) into the state:

\begin{equation}
|\psi^{-}\rangle=\frac{1}{\sqrt{2}}\left(|S\rangle_{1}|L\rangle_{2} - e^{i(\phi_{1}-\phi_{2})}|L\rangle_{1}|S\rangle_{2}\right)
\label{eq:SLvsLS}
\end{equation}

\noindent where the kets $|S\rangle$ and $|L\rangle$ represent, respectively, a single-photon in the short or long arms of the unbalanced MZ's, and $\phi_{1}$ and $\phi_{2}$ are the relevant phase differences (an overall phase-factor has been omitted).

When the path-length imbalances of the MZ's are set to match $\tau_{o}$, the two terms in eq. (\ref{eq:SLvsLS}) are indistinguishable, and lead to a coincidence counting rate between detectors $D_{1}$ and $D_{2}$ that goes like $R_{c} \sim 1+Cos(\phi_{1}-\phi_{2})$ \cite{liangPhD12}. This ``$SL$ vs. $LS$'' scenario is slightly different than typical ``$SS$ vs. $LL$'' Franson interferometery where the coincidence counting rate goes as $R_{c} \sim 1+Cos(\phi_{1}+\phi_{2})$ \cite{franson89}.  In both cases, the dependence on the difference or sum of spatially separated phases is the signature of nonlocal behavior. As in the more typical case, a visibility of this two-photon interference pattern greater than 71\% can be used to violate a Bell inequality using time-bin EPH's \cite{clauser78}.

\begin{figure}[t]
\includegraphics[width=3.5in]{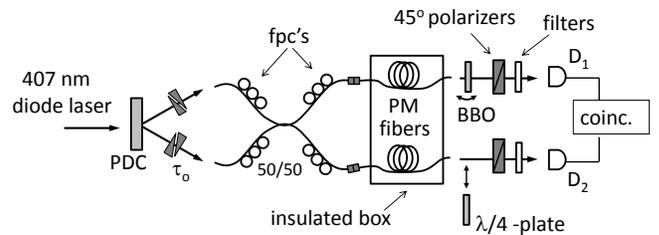}
\caption{Experimental apparatus used to implement the conceptual overview of Figure \protect\ref{fig:experimentoverview}. A PDC source provides the two background photons , and Polarization Maintaining (PM) fibers are used to implement the polarization-based MZ interferometers. The phases $\phi_{1}$ and $\phi_{2}$ are controlled by an adjustable birefringent BBO slab and a quarter-wave plate as described in the text. Fiber polarization controllers (fpc's) are used to define and maintain the relevant polarization states, and 10 nm bandpass interference filters are used to define a PDC photon coherence time of $\sim$ 200 fs.}
\label{fig:experimentalsetup}
\end{figure}

Figure \ref{fig:experimentalsetup} shows our experimental setup used to implement the idea of Figure \ref{fig:experimentoverview}. A type-I PDC source (0.7 mm thick BBO crystal) was pumped by a cw ultraviolet diode laser ($\sim$ 40 mW at 407 nm, $\sim$ 0.1 nm linewidth). The source produced horizontally polarized photon pairs at 814 nm which were coupled into a 50/50 single-mode-fiber beamsplitter. Translatable glass wedges were used to introduce and adjust the time-bin separation $\tau_{o}$, and fiber polarization controllers (fpc's) on the input ports of the beamsplitter were used to prepare the states $|H\rangle$ and $|V\rangle$.

The polarization-based unbalanced MZ's of Figure \ref{fig:experimentoverview} were implemented using two Polarization-Maintaining (PM) fibers with their fast and slow axes oriented in the horizontal and vertical directions, respectively. The PM fibers were 2 meters long, which gave a MZ path length imbalance of $\tau_{o} \sim$ 2.6 ps for the 814 nm photons. This imbalance greatly exceeded the coherence-time of the PDC photons ($\sim$ 200 fs), as required in Franson interferometry \cite{franson89}.

The phase $\phi_{1}$ was controlled by adding a birefringent slab (in this case, a second 0.7 mm thick BBO crystal) with its fast and slow axes aligned to match those of the PM fiber. Twisting this slab around the vertical axis changed its effective thickness and allowed smooth scanning of $\phi_{1}$.  For completeness, we added or removed a quarter-wave plate at the output of PM fiber 2 for two different relative values of $\phi_{2}$ ($\frac{\pi}{2}$ or $0$).

From a technical point of view, this setup was very robust against phase drifts. Indeed, the two-photon state of eq. (\ref{eq:SLvsLS}) is insensitive to the relative phase difference between the two original photons \cite{liangPhD12}. Consequently, the delay $\tau_{o}$ introduced by the glass wedges (and set equal to that of the PM fibers) only needed to be adjusted to within the coherence length of the PDC photons, rather than interferometrically stabilized; this is similar to the case of typical Hong-Ou-Mandel experiments \cite{hong87}. Additionally, by placing the relatively short PM fibers in a thermally insulated box, temperature dependent phase-drifts in the MZ's were greatly reduced and active locking was not needed to stabilize  $\phi_{1}$ and $\phi_{2}$.

\begin{figure}[t]
\includegraphics[width=2.75in]{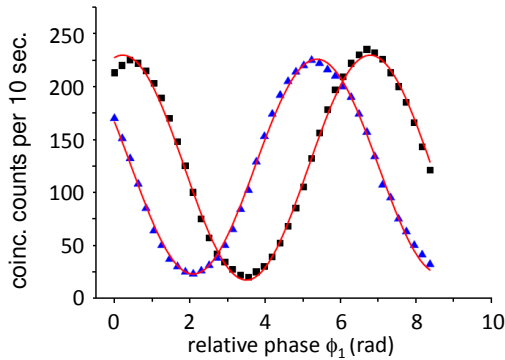}
\caption{(color online) Experimental violation of Bell's inequality with time-bin EPH's. The solid lines are least-squares fit to the data constrained by a common period. The visibility of the black-squares curve is ($86.1 \pm 0.$5)\%, and the blue-triangles curve is ($81.7 \pm 0.$5)\%.}
\label{fig:results}
\end{figure}

Figure \ref{fig:results} shows typical experimental results obtained with this setup. The data shows the expected sinusoidally varying coincidence counting rate as a function of $\phi_{1}$, for the two different relative values of $\phi_{2}$ ($0$ and $\frac{\pi}{2}$).  The MZ phase-difference $\phi_{1}$ could be scanned through $2\pi$ by a very small twist of the birefringent slab ($\sim 1^{o}$), so the phase was approximated as linearly changing with the twist-angle setting on a precision rotation stage. The calibration of the x-axis in the figure was verified by using the same scanning range and the known phase-shift of $\frac{\pi}{2}$ that exists between the two data sets (ie. with and without a known quarter-wave plate).

The two data sets in Figure \ref{fig:results} were best-fit to sinusoidal curves (constrained to a common period) with visibilities of ($86.1 \pm 0.$5)\% and ($81.7 \pm 0.$5)\%. The deviation from an ideal $100$\% visibility was most-likely due to standard mode-matching imperfections in PDC experiments, and we believe the slightly lower value for the second curve was introduced by non-ideal alignment of the additional quarter-wave plate. In any event, these high visibilities ($> 71$\% \cite{clauser78}) represent our proof-of-principle demonstration of a violation of Bell's inequalities using time-bin EPH's.

As a tangential side note, it is interesting to consider how the physical setup sketched in Figure \ref{fig:experimentoverview} can also be viewed as a time-bin analog of the original Shih-Alley experiment for generating polarization entanglement via post-selection \cite{shih88}. In the original Shih-Alley experiment, two orthogonal polarization qubits ($|H\rangle$ and $|V\rangle$) are mixed at a 50/50 beamsplitter, and measured with polarization qubit analyzers (ie. polarizers). In Figure \ref{fig:experimentoverview}, two orthogonal time-bin qubits ($|e\rangle$ and $|l\rangle$) are mixed at a 50/50 beamsplitter, and measured with time-bin qubit analyzers (ie. unbalanced MZ interferometers \cite{tittel00}). This analogy is particularly clear if one considers the setup of Figure \ref{fig:experimentoverview} with no polarization dependence.

\section{Summary and Discussion}
\label{sec:summary}

In this paper we have extended the original work on energy-time EPH's \cite{franson06,pittman06,afek10} to introduce the idea of time-bin EPH's. We have also considered the possibility of EPH's existing in backgrounds formed by single-photon states.

The initial proposal for energy-time EPH's considered cw coherent state background beams \cite{franson06}, whereas the initial experiments essentially considered a discretized version with the background spread out into infinitely long coherent pulse-trains \cite{pittman06,afek10}. Here we considered the case when the EPH's exist in a coherent-state background that is confined to two well-defined time-bins. When this coherent state background is further reduced to a single-photon background, the time-bin EPH state becomes physically equivalent to a particular time-bin entangled photon-pair state \cite{brendel99}. We used a conventional PDC photon-pair source to generate this background, and were thereby able to perform a proof-of-principle experimental violation of Bell's inequality using time-bin EPH's in a Franson-type interferometer \cite{franson89}.

The interpretation of any photon counting experiments related to EPH's is non-intuitive, as it is indeed the background photons which are being detected rather than any ``photon-holes''. In Franson interferometer experiments, the simplest explanation is that the existence of the EPH's suppresses the coincidence counting rate in a nonlocal way. Roughly speaking, the photon-holes prevent the possibility of two photons being found at the same time. This is particularly clear in the time-bin EPH states studied here, but applies to the energy-time case as well \cite{franson06}.

Methods to generate other forms of EPH's, such as those based on polarization or other degrees of freedom, remains an open question. The closely-related idea of generating entanglement through superpositions of correlated phase disturbances (rather than correlated loss) has recently been proposed \cite{kirby12}. We hope that states of this kind, as well as the time-bin EPH states considered here, provide new insight into the nature of entanglement and valuable resources for quantum information processing applications.

This work was supported by DARPA DSO under grant W31P4Q-10-1-0018.

{\em Note added: during preparation of this manuscript we became aware of a preprint \cite{martin12} which describes a telecom-band realization of the state in eq. (\ref{eq:ELvsLE}). The authors aptly named this the ``cross time-bin'' entangled state, and convincingly demonstrate its utility for quantum key distribution.}


\end{document}